\documentclass[16pt,doublespace]{aastex}
\begin{document}
\def\NL{$2{\times}10^{20}$cm$^{-2}$}
\def\NH{$N$(HI)}
\def\pkts{P_{KS}}
\def\eslash{$\rm \acute e$}
\def\etal{et al.}
\def\Lya{Ly$\alpha$ }
\def\lya{Ly$\alpha$ }
\def\smpy{M$_{\odot}{\rm \ yr^{-1} }$}
\def\smpykpc{M$_{\odot}{\rm \ yr^{-1} \ kpc^{-2}}$}
\def\smpympc{M$_{\odot}{\rm \ yr^{-1} \ Mpc^{-3}}$}
\def\Msolar{M$_{\odot}$}
\def\lcdm{$\Lambda$CDM }
\def\rvec{${\bf r}$}
\def\Rratio{${{\cal R}(v)}$}
\def\ndmp{17}
\def\nrun{8,500}
\def\kms{km~s$^{-1}$ }
\def\omgm{$\Omega_{{\rm M}}$}
\def\omgv{$\Omega_{\Lambda} $}
\def\omgas{$\Omega_{g}(z) $}
\def\dv{$\Delta v$}
\def\muG{$\mu$G}
\def\Bpar{$B_{\rm los}$}
\def\Vnu{$V({\nu})$}
\def\dvm{$\delta v$}
\def\vc{$V_{rot}(r)$}
\def\vrot{$V_{rot}(r)$}
\def\micron{$\mu$m}
\def\gamd{${\Gamma_{d}(\bf r)}$}
\def\gamdnr{${\Gamma_{d}}$}
\def\peq{$P_{eq}$}
\def\pmin{$P_{min}$}
\def\pmax{$P_{max}$}
\def\taunu{$\tau_{\nu}$}
\def\pgeom{$(P_{min}P_{max})^{1/2}$}
\def\kap{${\kappa ({\bf r})}$}
\def\dNdX{$d{\cal N}/dX$}
\def\kapnr{${\kappa}$}
\def\cm2{${\rm cm}^{-2}$}
\def\cm3{${\rm cm}^{-3}$}
\def\N#1{{N({\rm #1})}}
\def\f#1{{f_{\rm #1}}}
\def\rAA{{\rm \, \AA}}
\def\sci#1{{\rm \; \times \; 10^{#1}}}
\def\ltk{\left [ \,}
\def\ltp{\left ( \,}
\def\ltb{\left \{ \,}
\def\rtk{\, \right  ] }
\def\rtp{\, \right  ) }
\def\rtb{\, \right \} }
\def\ohf{{1 \over 2}}
\def\nohf{{-1 \over 2}}
\def\rhf{{3 \over 2}}
\def\smm{\sum\limits}
\def\perd{\;\;\; .}
\def\cmma{\;\;\; ,}
\def\semic{\;\;\; ;}
\def\sgint{\sigma_{int}}
\def\Nperp{$N_{0}$}
\def\frat{$f_{ratio}$}
\def\intl{\int\limits}
\def\rhodot{$\dot{\rho_{*}}$}
\def\rhodotz{$\dot{\rho_{*}}$$(z)$}
\def\rhodotz{$\dot{\rho_{*}}$$(z)$}
\def\rhosz{${\rho_{*}(z)}$}
\def\rhos{${\rho_{*}}$}
\def\und#1{{\rm \underline{#1}}}
\def\ps{$\dot{\psi_{*}}$}
\def\ms{$\dot{M_{*}}$}
\def\psav{$<$$\dot{\psi_{*}}$$>$}
\def\psavz{${<{{\dot{\psi_{*}}}}(z)>}$}
\def\ciis{C II$^{*}$}
\def\nh{$N_{\rm H I }$}
\def\Inu{$I({\nu})$}
\def\Inuc{$I_{c}({\nu})$}
\def\lc{$\ell_{c}$}
\def\lcav{$<l_{c}>$}
\def\lcr{$l_{cr}({\rm {\bf r}})$}
\def\lcrnr{$l_{cr}$}
\def\jnu{$J_{\nu}$}
\def\knu{$k_{\nu}$}
\def\junit{ergs cm$^{-2}$ s$^{-1}$ Hz$^{-1}$ sr$^{-1}$}
\def\lcunit{ergs  s$^{-1}$ per hydrogen atom}

\title{AN 84-{\muG} MAGNETIC FIELD IN A GALAXY AT
$Z$=0.692}

\author{ ARTHUR M. WOLFE\altaffilmark{1} and REGINA A. JORGENSON,\\ 
Department of Physics and Center for Astrophysics and Space Sciences; \\
University of California, San
Diego; \\
La Jolla, CA; 92093-0424\\
{\bf awolfe@ucsd.edu, regina@physics.ucsd.edu}}
 
\author{}

\author{ TIMOTHY ROBISHAW and CARL HEILES\\ 
Astronomy Department; \\
University of California; \\ 
Berkeley , CA; 94720-3411\\
{\bf robishaw@astro.berkeley.edu, heiles@astro.berkeley.edu }}

\author{and}

\author{JASON X. PROCHASKA\\ 
UCO-Lick Observatory; \\
University of California, Santa Cruz\\
Santa Cruz, CA; 95464\\
{\bf xavier@ucolick.org}}


{\bf The magnetic  field pervading our Galaxy is a crucial constituent of the 
interstellar medium:
it mediates the dynamics of interstellar clouds,
the energy density of cosmic rays, and the formation
of stars $^{1}$. 
The field associated
with ionized interstellar gas has been
determined through 
observations of
pulsars in our Galaxy. Radio-frequency measurements of
pulse
dispersion and the rotation of the plane of linear polarization, i.e.,
Faraday rotation,
yield an average value  {\bf \it B} $\approx$ 3 {\muG} (ref. 2).
The possible detection
of Faraday rotation of linearly polarized photons emitted by
high-redshift quasars $^{3}$ suggests 
similar magnetic fields are present in
foreground galaxies with redshifts $z$ $>$ 1.
As Faraday rotation alone, however, determines neither
the magnitude nor the redshift of the magnetic field, the strength of galactic
magnetic fields at redshifts $z$ $>$ 0 remains 
uncertain.
Here we report a measurement of a magnetic field of $B$ $\approx$ 84 {\muG}
in a galaxy at $z$ =0.692,
using  the same Zeeman-splitting technique that
revealed an average value of {\it B} 
= 6 {\muG} in the  neutral
interstellar gas of our Galaxy $^{4}$.
This is unexpected, as the leading theory
of magnetic field generation, the mean-field dynamo model, predicts
large-scale magnetic  fields to be weaker in the past rather than stronger. $^{5}$}

\begin{figure}
\figurenum{1}
\includegraphics[angle=0,width=.60\textwidth]{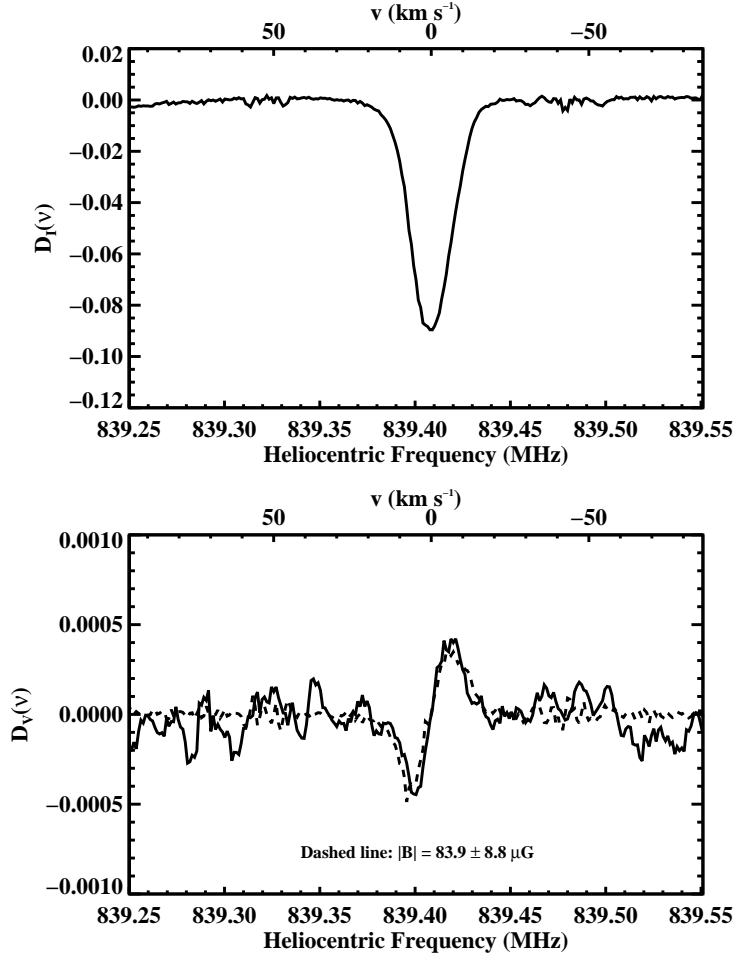}
\caption[]{\tiny {\bf Line-depth spectra of Stokes-parameters.} 
Data acquired in 12.6 hrs. of on-source integration
with the 
GBT radio antenna. Because the GBT
feeds detect only orthogonal linearly polarized signals,
while Zeeman splitting requires measuring 
circular polarization
to construct Stokes $V({\nu})$,
we generated $V({\nu})$ by cross-correlation 
techniques $^{23}$. 
The velocity $v$=0 {\kms} corresponds to
$z$=0.6921526. {\bf a,}
Line-depth function
${D_{I}}({\nu})$ $\equiv$ [{\Inu}$-${\Inuc}]/{\Inuc},
where  
$I({\nu})$$\equiv$
$s_{0}+s_{90}$, with
$s_{\theta}$ the power measured in linear polarization
position angle $\theta$, 
corresponds
to the total intensity spectrum, and  
{\Inuc} is a model fit to the {\Inu} continuum.
${D_{I}}({\nu})$ 
=exp[$-{\tau({\nu})}$]$-$1 where
$\tau({\nu}){\equiv}[{\tau({\nu})}_{0}+{\tau({\nu})}_{90}]/2$
is the average optical depth in the two
orthogonal states of linear polarization $^{4}$.
{\bf b,} Line-depth  function
${D_{V}}({\nu})$$\equiv$ 
$V({\nu})$/$I_{c}({\nu})$,
where 
{\Vnu}$\equiv$$s_{\rm RCP}$$-$$s_{\rm LCP}$ 
is difference in power between the
right-hand and left-hand circularly polarized (respectively
RCP and LCP) signals. 
Here ${D_{V}}({\nu})
=-[\tau_{V}({\nu})/2]{\rm exp}[-\tau({\nu})]$,
where 
$\tau_{V}({\nu}){\equiv}\tau_{\rm RCP}({\nu})-
\tau_{\rm LCP}({\nu})$ $<<$1 (ref. 4) is the 
difference between the optical depths of RCP and LCP
photons. 
For Zeeman splitting of the 21 cm line, the degeneracy of the
$F=0$ to $F=1$ hyperfine transition is removed since the 
$m_{F}=-1,0,+1$ states differ in energy. This results
in a small frequency difference between absorbed
LCP photons ($m_{F}=-1$) and RCP photons ($m_{F}=+1$).
{\Vnu}
is crucial for detecting Zeeman splitting because the orthogonal,
circularly polarized states of the photon are eigenstates of the 
spin angular momentum operator with eigenvalues $\pm$ $\hbar$,
that is, angular momenta directed along or opposite to the
direction of photon propagation.$^{24}$
When $B_{\rm los}=B$, transitions between
the hyperfine $F$
= 0   and $F$= 1  states occur
exclusively
through absorption of LCP or RCP photons through excitation
of the $m_{\rm F}$ = $-$ 1 and  
$m_{\rm F}$ = $+$ 1 hyperfine states respectively.
Because $V({\nu})$ is the difference in the RCP and LCP
intensities, the resulting $V({\nu})$ line profile
is the difference between two Gaussian absorption profiles
with frequency centroids shifted by $\Delta \nu_{B}$
=2.8($B_{\rm los}$/{\muG})$(1+z)^{-1}$ Hz
(where $B_{\rm los}$ is measured in microgauss).
The `S curve' is due to
the sign flip in RCP-minus-LCP intensity difference as $\nu$ passes
through line centre.} 
\label{fig:stokes_3C286}
\end{figure}

We detected Zeeman splitting of the
$z$=0.692, 21 cm absorption line in the direction of the quasar 3C 286
(refs. 6,7) using the 100-m Robert C. Byrd Green Bank Telescope
(GBT) of the National Radio Astronomy Observatory.
The absorption
arises in a damped {\lya} system (henceforth denoted DLA-3C286)
that is drawn from 
a population of neutral
gas layers widely thought to be the progenitors of modern
galaxies $^{8}$.
The radio data for DLA-3C286 are summarized in  
Fig.~{\ref{fig:stokes_3C286}}, which shows the line-depth
spectra constructed from   
the measurable quantities used to describe polarized radiation,
that is, the Stokes parameters.
We show 
the line-depth spectra constructed from the
$I({\nu})$  and  $V({\nu})$ Stokes parameters
(where $\nu$ denotes frequency) near the 839.4 MHz frequency centroid
of the redshifted 21 cm absorption line.
Fig.~{\ref{fig:stokes_3C286}}a shows the line-depth spectrum
constructed from 
$I({\nu})$.
A 
Gaussian fit to the absorption
line in 
Fig.~{\ref{fig:stokes_3C286}}a
yields a redshift $z$=0.6921526$\pm$0.0000008, central optical depth
of
$\tau_{0}$=0.095$\pm$0.006,
and a velocity dispersion $\sigma_{v}$=3.75$\pm$0.20 {\kms}, which are in
good agreement with 
previous results $^{6,7}$.

In Fig.~{\ref{fig:stokes_3C286}}b we plot the line-depth spectrum
constructed from $V({\nu})$, which shows
the classic
`S curve' pattern expected for Zeeman splitting. 
From our least squares fit to the data,
we find 
{\Bpar}=83.9{$\pm$}8.8 {\muG}, 
where 
{\Bpar} is the magnetic field component projected along the
line of sight 
(we note that the direction of {\Bpar} is
unknown since the instrumental sense of circular
polarization was not calibrated).
This magnetic field differs in two respects 
from the magnetic fields obtained from Zeeman splitting
arising in interstellar clouds in the Galaxy.  
First, the field strength corresponds to
the line-of-sight component of the  mean
field $<B_{\rm los}>$
averaged
over transverse dimensions exceeding 
200 pc, as very-long-baseline interferometry
observations of the 21-cm absorption line show
that the gas layer must must extend across
more than 0.03 {\arcsec} to explain
the difference between the velocity centroids of the fringe amplitude
and phase-shift spectra $^{9}$ (although the data are consistent
with a $B$-field coherence length less than 200 pc,
the resulting gradient in magnetic pressure would produce velocity
differences exceeding the shift of $\approx$ 3 {\kms}  across
200 pc detected
by very-long-baseline interferometry). By contrast,
the transverse dimensions of radio beams subtended
at neutral interstellar clouds in the Galaxy are typically less than 1 pc.
Second, this field is at least an order of magnitude stronger than
the 6-{\muG} average of $B$ fields inferred from Zeeman splitting for
such clouds $^{4}$.

\begin{figure}
\figurenum{2}
\includegraphics[angle=0,width=.80\textwidth]{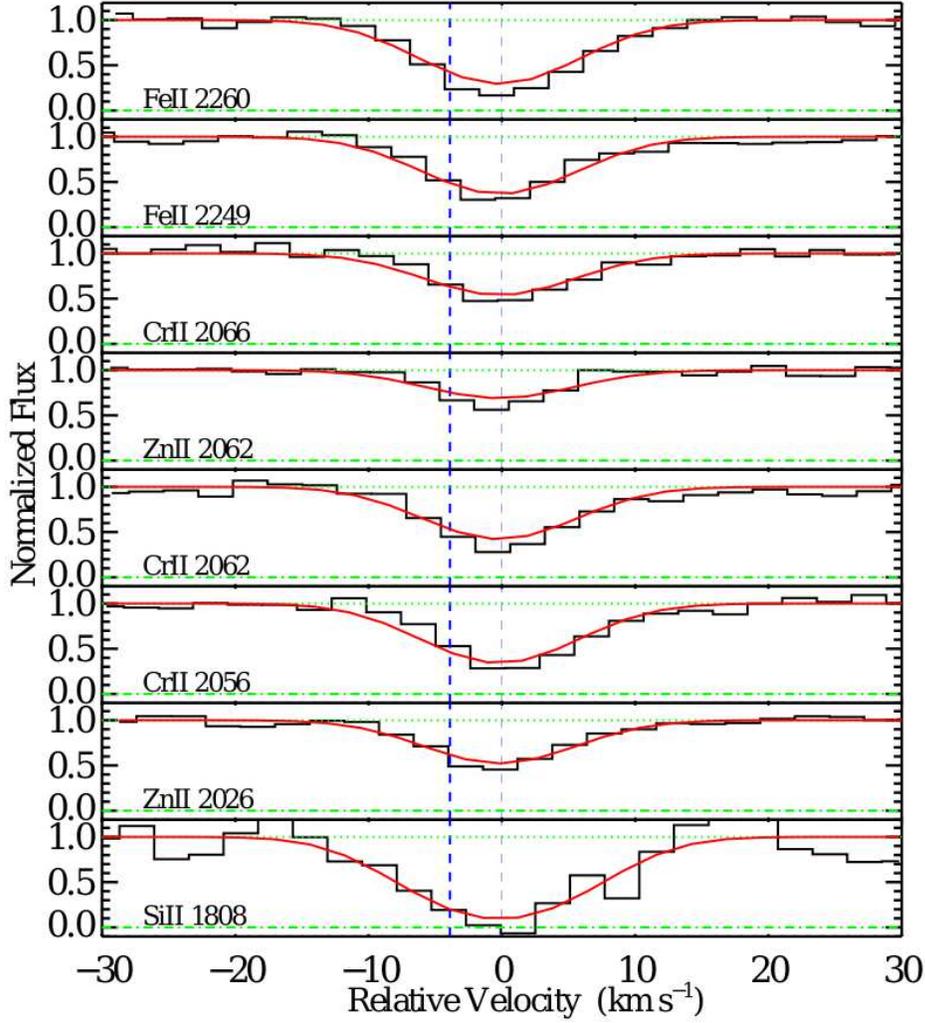}
\caption[]{\tiny {\bf HIRES velocity profiles for dominant low-ionization
states of abundant elements
in the 21 cm absorber towards 3C 286.} Spectral
resolution $\Delta v$=7.0 {\kms} and the average signal-to-noise
ratio per 2.1 {\kms} pixel is about 30:1.
The  bold dashed vertical line
denotes the velocity centroid of single-dish 21 cm absorption
feature and the faint dashed vertical line denotes
the velocity centroid of the resonance lines shown in the
figure. Our least squares fit of Voigt profiles (red) to the data (black)
yields
ionic column densities as well as the redshift centroid and velocity
dispersion shown in Table 1 (lower and upper
green horizontal lines refer to zero and unit normalized fluxes).
Because refractory
elements such as Fe and Cr
can be depleted onto dust grains $^{25}$, we used the volatile elements 
Si and Zn to derive a logarithmic  metal
abundance with respect to solar abundances of
[M/H]=$-$1.30. 
The
depletion ratios [Fe/Si] and [Cr/Zn] were then used to derive
a conservative upper limit on the logarithmic dust-to-gas ratio
relative to Galactic values
of [D/G] $<$$-$1.8.}
\label{fig:hires}
\end{figure}

We obtained further information about conditions in the absorbing gas
in DLA-3C286
from
accurate optical spectra acquired
with the HIRES Echelle spectrograph on the Keck I 10 m
telescope.
Fig.~{\ref{fig:hires}} shows velocity profiles for
several
resonance absorption lines arising from dominant
low-ionization states of abundant elements. 
The results of our least squares fit
of  Voigt profiles to the data  
are
shown in Table 1, where the optical redshift is displaced
$+$3.8 $\pm$0.2{\kms} from the 21 cm redshift.
This solution also yields
ionic column
densities from which we derived the logarithmic metal abundances 
with respect to solar abundances, [M/H], and 
dust-to-gas ratios with respect to the Galactic interstellar medium, [D/G]. 
These are among the lowest values
of [M/H] and [D/G] deduced
for damped {\lya} systems at $z$=0.7 (refs. 10, 11).  The 
low metallicity indicates a history of low
star formation rates (SFR). Because
the intensity of far ultra-violet  radiation emitted by young massive stars
is proportional to the concurrent SFR per unit area $\Sigma_{\rm SFR}$,
low values of $\Sigma_{\rm SFR}$ should result in low grain photoelectric
heating rates per H atom, $\Gamma_{\rm pe}$ (ref. 11).
This is
consistent with the low upper limit, 
$\Gamma_{\rm pe}$ $<$ 10$^{-27.4}$ {\lcunit},  obtained
by combining 
the assumption of
thermal balance with the absence of 
{\ciis}  absorption (that is, absorption from C II in
the excited $^{2}P_{3/2}$ fine structure state) at a wavelength
of 1335.7 {\AA} in the
previous low-resolution Hubble Space Telescope (HST) spectra of 3C 286 
(ref. 12), and indicates 
$\Sigma_{\rm SFR}$ $<$ 10$^{-2.9}$ {\smpykpc} (95 $\%$ confidence level), 
which is less 
than the solar-neighborhood
value of  
10$^{-2.4}$ {\smpykpc} (ref. 13).

As a result, we have detected an unusually strong 
magnetic field at $z$ = 0.692 with a coherence length 
that probably exceeds 200 pc
in neutral gas that is quiescent, metal-poor, nearly
dust-free, and presents little evidence
for star formation. 
To model this configuration,
we first
consider the magnetostatic equilibrium of a plane-parallel
sheet with in-plane magnetic field $B_{\rm plane}$ orthogonal to
the vertical gravitational field exerted by gas with
perpendicular mass surface density, $\Sigma$. 
In magnetostatic equilibrium
the total midplane pressure,
($B_{\rm plane}^{2}/8{\pi})  + {\rho}{\sigma}_{v}^{2}$, 
equals the `weight' of the
gas, ${\pi}G{\Sigma}^{2}/2$, where $\rho$ is the mass volume
density of the gas and $G$ is the gravitational constant. 
However, because the pressure-to-weight ratio
exceeds 715 in DLA-3C286, 
the magnetized gas cannot be confined by its self-gravity.
Therefore, self-consistent magnetostatic configurations are
ruled out unless the contribution of stars to 
$\Sigma$ exceeds $\approx$ 350 {\Msolar}pc$^{-2}$. Although 
this is larger than the 
50 {\Msolar}pc$^{-2}$
surface density perpendicular to the solar neighborhood,
such surface densities are common in the central 
regions of galaxies.  In fact high surface densities
of stars
probably
confine the highly magnetized gas in the nuclear rings
of barred spirals. These exhibit {\it total} field strengths
of $\sim$ 100 {\muG}, inferred by 
assuming equipartition of magnetic and cosmic-ray
energy densities$^{1}$.
However, because the rings are
associated with regions of active star formation, 
high molecular content, and
high dust content,
they are unlikely sites of the magnetic field detected
in DLA-3C286.

 On the other hand the absorption site might consist
of highly magnetized gas confined by the gravity
exerted by a disk of old stars.
The  H I disks found at the centers of early-type
S0 and elliptical galaxies $^{14}$
are possible prototypes. 
Support for this idea
stems from
a high-resolution image obtained with the Hubble Space Telescope:
a Wide Field and Planetary Camera 2 (WFPC2) 
$I$-band image, from which the quasar has been 
subtracted, reveals residual emission spread over angular 
scales of $\sim$ 1 {\arcsec} (ref. 15). 
The asymmetry of the light
distribution with respect to the  point-source quasar 
suggests that some of the light is emitted by a foreground
galaxy with a brightness centroid displaced  less
than 0.5 {\arcsec} from the quasar. The location
of diffuse emission in the direction of an amorphous
object detected 2.5 {\arcsec} from
the quasar in ground-based imaging $^{16}$ further suggests that
the diffuse emission comes from  central regions
of the amorphous object. A recent reanalysis of the WFPC2
image shows the amorphous object to be a filament resembling a
spiral arm or  tidal tail (H.-W. Chen,  personal communication), 
that is, the outer appendage 
of a
galaxy centered within  a few kpc of the quasar sightline.

However, the magnetic field detected in DLA-3C286
may not be confined by gravity in an equilibrium configuration.
Rather, the detected field may be enhanced by a shock (F.H.Shu, personal 
communication).
Assuming a typical value of $B_{\rm plane}$$\approx$ 5 {\muG}
for the equilibrium field of the preshock gas, we find that a shock-front
velocity of $\approx$ 250 {\kms} will result in a post-shock field 
strength of $\approx$ 100 {\muG} in the limit of flux freezing
in a radiative shock with post-shock density 
of $\approx$ 10 cm$^{-3}$. This scenario seems plausible because 250 {\kms}
is reasonable for the impact
velocity generated by the merger between the gaseous disks
of two late-type galaxies, and the WFPC2 image is consistent
with the presence of two foreground galaxies. 
But
the second disk would
create another  set of absorption lines displaced  $\ge$
250 {\kms} from the redshift of DLA-3C286, which
is the only redshift observed.
By contrast, the merger between a gaseous disk and an elliptical
galaxy could result in only
one damped {\lya} system redshift, as a significant fraction of 
ellipticals do not contain  H I disks $^{14}$. In this case a  shock front
moving in the plane of the disk galaxy
would be generated by 
the gravitational impulse induced by 
the elliptical moving normal to the plane. 
Preliminary estimates indicate that an elliptical with
a modest mass, $M$=2{$\times$}10$^{11}$ {\Msolar}, and
impact velocity of $\approx$ 300 {\kms} would produce
a cylindrical shock of sufficient strength to boost an
initial field with $B_{\rm plane}$ $\approx$ 10 {\muG}
to $\approx$ 100 {\muG}.

Let us examine these scenarios more closely. 
The quiescent
velocity field of the gas fits in naturally with
the `magnetostatic equilibrium' scenario, because
the low
value of $\Sigma_{\rm SFR}$ suggests a low rate of
energy injection into the gas by supernovae $^{17}$, which could
result in a velocity dispersion of 
$\sigma_{v}$ $\approx$ 4 {\kms}. Moreover, the 
weak radio jets associated with early-type galaxies
containing central H I disks are natural sources of $B$ fields
for these disks. However, 21 cm absorption
measurements of such disks in nearby galaxies reveal
the presence of absorption-line widths far broader than
the narrow-line width of DLA-3C286 (ref. 18). Also,
it is unclear  whether or not the
high surface density of old stars required to confine the $B$
fields are present in these disks, and  whether or not the build-up of
$B_{\rm plane}$ to 100 {\muG} is possible in the 4 to 5 Gyr
age of the disk. In the `merger scenario' the
dynamo need only build up to $\approx$ 10 {\muG} in
the same time interval, but it is then necessary to   
explain why the post-shock velocity field averaged
over length scales of 200 pc is so quiescent.
Furthermore, the probability $p$ for detecting
$\sim$ 100-{\muG} magnetic fields in a random sample of 21 cm
absorbers is small. Our estimates, based on the
merger fraction of galaxies with $z$ $\sim$ 1 (ref. 19) and on the duration
time for magnetic  field enhancement, suggest that $p$ $\approx$ 0.005 to 0.03:
either we were lucky, or some characteristic of DLA-3C286,
such as narrow line width, is a signature of 
strong magnetic fields. 

Therefore, it is premature to decide among
these and other possible models to explain the presence
of the 84 {\muG} magnetic field in DLA-3C286. However,
our data  support
the inference from recent tenetative
evidence for Faraday rotation in high-$z$ quasars $^{20}$ 
that
magnetic fields are generic features of galaxies at high
redshifts,  which potentially have a more important role
in galaxy formation and evolution $^{21}$ than hitherto realized.
Specifically, the highly magnetized gas that we have
detected could suppress gravitational collapse and hence
may be a reason for the low {\it in situ} star formation rates of 
high-$z$ \\ DLAs $^{22}$. We plan to test this hypothesis by
using the GBT to search for Zeeman
splitting in high-redshift DLAs exhibiting 21 cm absorption.

\clearpage


\acknowledgements

{\bf Acknowledgements} 
We wish to thank F. H. Shu for suggesting the merger model
and H.-W. Chen for providing us with her
reanalysed images of 3C 286..
We also thank F. H. Shu, E. Gawiser, and A. Lazarian
for comments and the US National Science Foundation for financial
support. The GBT is one of the facilities of the 
National Radio Astronomy
Observatory, which is a center of the National Science Foundation
operated under cooperative agreement by Associated Observatories, 
Inc. A. M. W., R. A. J., and J. X. P. are Visiting
Astronomers at the W. M. Keck Telescope. The Keck Oservatory
is a joint facility of the University of California, the
California Institute of Technology and the National Aeronautics
and Space Administration.


\begin{table} 
\begin{center}
\begin{tabular}{lcc}
\multicolumn{3}{c}{Redshift $z$=0.69217485{$\pm$}0.00000058}\\
\cline{1-3}
\multicolumn{3}{c}{Velocity Dispersion $\sigma_{v}$=3.08$\pm$0.13 \ {\kms}}\\
\cline{1-3}
Ion $X$ &log$_{10}$$N(X)$&[X/H]   \\
   &cm$^{-2}$&    \\
\tableline
H I & 21.25${\pm}$0.02&---  \\
Fe II  & 15.09${\pm}$0.01&$-$1.66{$\pm$}0.02  \\
Cr II  & 13.44${\pm}$0.01&$-$1.48{$\pm$}0.02  \\
Zn II  & 12.53${\pm}$0.03&$-$1.39{$\pm$}0.03  \\
Si II  & $>$ 15.48&$>$$-$1.31  \\
\end{tabular}
\end{center}
\caption{{\bf Physical Parameters of DLA-3C286 Inferred 
from Optical Absorption}} \label{data}
\end{table}


\newpage

\newpage

\end{document}